\documentclass[12pt]{article}
\usepackage{amsfonts,amsmath,epsfig,graphicx,amssymb}
\usepackage{amsmath,epsfig,graphicx,amssymb}
\usepackage[margin=1in]{geometry}
\usepackage{amsmath}
\usepackage{amsfonts}
\usepackage{amssymb}
\usepackage{makeidx}
\usepackage{graphicx}
\makeatletter
\DeclareRobustCommand{\cev}[1]{%
  \mathpalette\do@cev{#1}%
}
\newcommand{\do@cev}[2]{%
  \fix@cev{#1}{+}%
  \reflectbox{$\m@th#1\vec{\reflectbox{$\fix@cev{#1}{-}\m@th#1#2\fix@cev{#1}{+}$}}$}%
  \fix@cev{#1}{-}%
}
\newcommand{\fix@cev}[2]{%
  \ifx#1\displaystyle
    \mkern#23mu
  \else
    \ifx#1\textstyle
      \mkern#23mu
    \else
      \ifx#1\scriptstyle
        \mkern#22mu
      \else
        \mkern#22mu
      \fi
    \fi
  \fi
}

\makeatother
\usepackage{appendix}
\newcommand{\beq}{\begin{equation}}
\newcommand{\eeq}{\end{equation}}
\newcommand{\ben}{\begin{eqnarray}}
\newcommand{\een}{\end{eqnarray}}
\date{}
\begin{document}

\title{The hidden Lorentz Covariance of Quantum Mechanics }
\author{Partha Nandi\footnote{pnandi@sun.ac.za}\\and\\
Frederik G. Scholtz\footnote{fgs@sun.ac.za} \\
Department of Physics,\\
University of Stellenbosch, Stellenbosch-7600, South Africa}

\maketitle

\begin{abstract}
This paper introduces a systematic algorithm for deriving a new unitary representation of the Lorentz algebra ($so(1,3)$) and an irreducible unitary representation of the extended (anti) de-Sitter algebra ($so(2,4)$) on $\mathcal{L}^{2}(\mathcal{R}^{3},\frac{1}{r})$.  This representation is equivalent to a representation on $\mathcal{L}^{2}(\mathcal{R}^{3})$, and the corresponding similarity transformation is identified.  An explicit representation in terms of differential operators is given, and it is shown that the inner product is Lorentz invariant. Ensuring Lorentz covariance demands a modification of the Heisenberg algebra, recognized as a phase space algebra at the interface of gravitational and quantum realms (IGQR), which we consider subordinate to Lorentz covariance. It is also demonstrated that time evolution can be cast in a manifestly covariant form.  Each mass sector of the Hilbert space carries a representation of the Lorentz algebra, and the (anti) de-Sitter algebra on each mass sector contracts to the Poincare algebra in the flat configuration and momentum space limits.  Finally, we show that three-dimensional fuzzy space also carries a unitary representation of these algebras, algebraically equivalent to the $\mathcal{L}^{2}(\mathcal{R}^{3},\frac{1}{r})$ representation but not necessarily equivalent as representations.  Several outstanding issues are identified for future exploration.

\end{abstract}

\maketitle

\section{Introduction}
\label{Introduction}

Quantum mechanics and general relativity (GR) stand as the twin pillars of our contemporary understanding of the physical world. The classical theory of gravity, rooted in deterministic solutions to the Einstein field equations, employs a commutative formalism and ensures causal time evolution. It excels in elucidating observations like gravitational waves on cosmic and galactic scales. In contrast, quantum mechanics thrives in describing the quantum world at atomic scales. It introduces non-determinism in quantum phenomena, operates under a non-commutative formalism, and mandates instantaneous collapses during quantum measurements \cite{Zu}. However, at very short time and length scales of the order of the Planck time and length, these two theories seem to be incompatible, and most likely both are not the correct description of the world at these scales.  Instead, a new, as yet unknown, theory that unites both of these in a consistent framework seems to be required.  This new framework is commonly referred to as quantum gravity. Clearly, quantum mechanics and GR must be good approximations of quantum gravity at the appropriate length scales.\\

Despite many attempts over the past century, quantum gravity is still alluding us. Indeed, the absence of a quantum theory of gravity may be the reason why we struggle to explain such puzzling phenomena as dark matter, dark energy, and the cosmological constant problem \cite{Zuk, l}.  It may even be at the root of two closely related fundamental problems in quantum mechanics, namely, measurement and quantum-to-classical transition problems \cite{l1,l2,l3,k3}.  At this stage, there also does not seem to be a convergence of opinion.  String theory, our most promising candidate for quantum gravity \cite{k4}, is faced with several serious conceptual problems \cite{l4}.  At the other extreme, there is even some scepticism about whether gravity is truly quantum or not \cite{l6}.  Indeed, it is not even known whether a gravitational field may exist in a quantum superposition of states \cite{l5}. Clearly, we need observational data to guide us at this point, which may come from the recent and rapid improvement in observational precision that may reveal phenomenological signals of quantum gravity \cite{l6,l7}.\\

Conceptually, the quantization of general relativity (GR) can be approached in two main ways. One method involves quantizing the gravitational field itself, represented by gravitons, metric fluctuations, or their extensions. Alternatively, quantization can be applied to the underlying space-time geometry. Since GR portrays gravity as the curvature of space-time, these approaches are interconnected. Various attempts to address quantum gravity highlight a consistent theme: the need for a fundamental reevaluation of space-time at small length scales, often referred to as quantum space-time.
Despite the multitude of approaches, the quest for quantum space-time lacks clear guiding principles. Nevertheless, a few indicative directions have emerged in the pursuit of understanding the nature of quantum space-time. The first is the realization, emphasised by Doplicher et al \cite{l8}, that it is operationally impossible to localize an event with arbitrary accuracy without encountering a gravitational instability, which provides compelling arguments for non-commutative space-time as a candidate for quantum space-time. The second is Lorentz covariance.  The Lorentz covariance of physical laws is a principle deeply embedded in relativity and strongly supported by observation. Sacrificing it may have unwanted observational consequences. From these two perspectives, the ideal would be a non-commutative, Lorentz covariant space-time, and quantum theory \cite{l9}.
Further compelling arguments in favour of non-commutative space-time as a candidate for quantum space-time arise from GR and Born's reciprocity argument \cite{l10}.  In GR, it is commonly accepted that space is curved, which implies non-commutative momentum.  Thus, the notion of non-commutative momentum is part and parcel of GR.  The only reason why space is treated as commutative in GR is because momentum space is not viewed as curved.  However, if one adopts Born's reciprocity argument, which states that there is a position-momentum duality of the form $\hat{x}\rightarrow \hat{p}$ and $\hat{p}\rightarrow -\hat{x}$ that leaves the Heisenberg commutation relation invariant, one would also expect curvature in momentum space if there is curvature in space, and thus also non-commutativity of coordinates. This implies a profound connection between curvature and non-commutativity, suggesting that they are two facets of a unified underlying reality. This relationship between curved momentum space and non-commutative geometry was recently explored in \cite{11,11+,11-}. \\

The realization that quantum space-time may be non-commutative drove several developments in the recent past, most notably non-commutative geometry \cite{12,13} and non-commutative quantum field theory \cite{14}. The significant motivation behind these developments is the prospect of non-commutativity serving as a distinctive feature of quantum gravity \cite{14-}. For a comprehensive overview, refer to the recent review on non-commutative field theory and gauge theory \cite{14+}. Recent studies have also demonstrated that a Moyal-like non-commutative space-time geometry naturally emerges due to the perturbative aspects of quantum gravity \cite{14++}.  However, the most naive model of non-commutative space-time using constant commutation relations breaks Lorentz covariance \cite{12++}, and a deformation of the Lorentz algebra is required to restore it \cite{14+++,13+++,RB}.  An unwanted consequence is the mixing of short and long length scales, commonly referred to as ultraviolet-infrared mixing \cite{15}. \\

On a historical note, the introduction of non-commutative space-time is a very old concept that was introduced by Snyder\cite{16} to regularize the divergence in quantum field theory. This uses a non-commutative model of space-time that does not violate Lorentz covariance. Following that, the Snyder model was extended to account for a de Siiter background in \cite{17}. This generalization was motivated by the necessity of including a cosmological constant ($\Lambda$), among the bare parameters of a theory of quantum gravity.  This clearly indicates that the Heisenberg uncertainty relation should be modified by quantum gravity corrections, which are very much reminiscent of the Extended Generalized Uncertainty Principle (EGUP) \cite{18}. However,  the commutation relations of the translation generators with the position variables depend on the specific choice of coordinates on the hyperboloid of the Snyder de-Sitter background. Usually, the most natural description of the Snyder de Sitter phase space algebra can be expressed in terms of operators acting on six-dimensional Minkowski space time \cite{19}.\\ 

Recent findings hint at a potential reconciliation between the concept of a 4D flat Lorentzian space-time and the principles of quantum mechanics, especially in the domain of harmonic oscillators \cite{20+}. This challenges the conventional belief in space-time as a fundamental concept, aligning with the viewpoint advocated by physicists like Pauli \cite{20-}. This raises a fundamental question: Can quantum mechanics and the Lorentz covariance of physical laws coexist harmoniously?\\

The aforementioned general observations lay the groundwork for the present paper. As may have become evident, our main aim here is to present a systematic approach to formulate a fully Lorentz covariant version of quantum mechanics applicable to both commutative and non-commutative spaces and to investigate the relationships between them. At first glance, this endeavor might appear challenging. The traditional view of Lorentz covariance requires space and time to be treated on the same footing, yet in quantum mechanics we construct the Hilbert space at a fixed time slice, and subsequently the inner product is not Lorentz invariant \cite{20}. Furthermore, time and space are treated on different footings in the Schr\"odinger equation \cite{20++}.  These were indeed the main issues faced in the first attempts to formulate relativistic quantum mechanics on commutative space.  This forced Dirac to replace the Schr\"odinger equation with a covariant equation of motion, the famous Dirac equation.  The issues regarding the construction of the Hilbert space can be resolved in a second quantized or quantum field theoretic setting.

Fortunately, it turns out that Lorentz covariance does not require the treatment of space and time on the same footing.  This has clearly not been appreciated in the literature and is one of the take-away messages of this paper.  The trade-off to achieve this is precisely the necessity of the deformation of the Heisenberg algebra (notably, it has been previously noted that relativistic corrections to quantum mechanics, derived from the single-particle sector of scalar quantum field theory, may lead to a deformation of the Heisenberg algebra \cite{21}), rather than the Lorentz algebra.

There are two routes to the construction of Lorentz-covariant quantum mechanics.  The first is to construct non-commutative quantum mechanics on three-dimensional fuzzy space, as was done in \cite{22,23}.  This construction exhibits puzzling features of a Lorentz covariant theory \cite{25}, which suggests the possibility that it may be a Lorentz covariant theory.  A closer inspection of this theory indeed reveals that the Hilbert space of this quantum theory carries a representation of the Lorentz $so(1,3)$ and, indeed, the (anti) de Sitter algebra $so(2,4)$.  We briefly review this point of view in Section 6. This approach is rather ad hoc, as it is based on a seemingly ad hoc choice of quantum space-time.  It will be much more gratifying if we can show directly that this is indeed the quantum space-time that emerges from the construction of a Lorentz covariant quantum theory.  This is the second route and the one we prefer to follow here.  These two routes lead to completely equivalent algebraic structures.\\

Many phenomenological consequences of non-commutative quantum mechanics in three-dimensional fuzzy space have already been worked out \cite{26,27,30,31}. Even the consequences at large length scales, which indicate a modification of Newtonian dynamics, have been explored \cite{32}. These are also the most likely observable phenomena, given the recent advancements in observational astronomy.  The algebraic equivalence between the current construction and non-commutative quantum mechanics on three-dimensional fuzzy space suggests that those conclusions may also hold here.  However, a more careful analysis is required here as the two representations are seemingly non-equivalent, but this falls outside the scope of the current paper.\\

The ultimate aim of these endeavors is to find clear phenomenological signals that can support or falsify the proposed structure of quantum space-time.  As the construction proposed here is very tight, suggesting that it may even be unique, this may provide us with key insights into the potential structure of a theory of quantum gravity. \\

The paper is organized as follows: In section \ref{Lorentz}, we construct a representation of the Lorentz algebra on $\mathcal{L}^{2}(\mathcal{R}^{3},\frac{1}{r})$ through a Hopf fibration and show its equivalence to a representation on $\mathcal{L}^{2}(\mathcal{R}^{3})$. In section \ref{dynamics}, we show that the Schr\"odinger equation can be cast in a covariant form.  In section \ref{Fuzzy}, we briefly review the construction of non-commutative quantum mechanics on three-dimensional fuzzy space and show its algebraic equivalence with the construction on $\mathcal{L}^{2}(\mathcal{R}^{3})$ given in section \ref{Lorentz}. In section \ref{Conclusions}, we summarize our results and draw conclusions.  Some intermediate results are collected in an appendix.\\

\section{Representation of the Lorentz and (anti) de Sitter algebras on $\mathcal{L}^{2}(\mathcal{R}^{3})$ }
\label{Lorentz}
The key to the construction of covariant quantum mechanics outlined in the introduction, is the construction of a representation of the Lorentz and (anti) de Sitter algebras on $\mathcal{L}^{2}(\mathcal{R}^{3})$, i.e. the space of square integrable functions on $\mathcal{R}^{3}$.  As will immediately be noted, time does not appear in this representation, which is at the origin of the claim that it is possible to have Lorentz covariance without treating space and time on the same footing.  The construction is done in a very systematic way, exploiting a Hopf fibration \cite{32++}.

\subsection{Spinor representation of the Lorentz and (anti) de Sitter algebras}
\label{Spinor}
Consider a dimensionless complex doublet $Z=(z_{1}~z_{2})^{T}\in \mathcal{C}^{2}$ which transforms as a spinor under $SU(2)$ transformations and is normalised as follows:

\begin{equation}
   Z^{\dagger}Z= r.
\end{equation}

This clearly suggests that the spheres $\mathcal{S}^{3}$ have been foliated to $\mathcal{R}^{4}$ with radii $\sqrt{r}$.

The embedding $\mathcal{S}^{2}$ with radius $r$ is specified by the dimensionless Cartesian coordinates of $\mathcal{R}^{3}$ and given by

\begin{equation}
  x_{1}=r \sin\theta \cos\phi,\;x_{2}=r \sin\theta \sin\phi,\;x_{3}=r \cos\theta.
\end{equation}

 Based on the Hopf fibration these Cartesian coordinates are given by the $\mathcal{C}^{2}$ coordinates as
 \begin{equation}
 x_{i}= Z^{\dagger} \sigma_{i} Z=(\sigma_{i})_{\alpha\beta}\bar{z}_{\alpha}z_{\beta};~~~~~~~~~ i=1,2,3, ~and ~\alpha,\beta=1,2.
 \end{equation}
 with $\sum_ix^{2}_{i}=r^{2}$.  This allows to parameterise the complex doublets above as follows:
\begin{equation}
	\begin{pmatrix}
	z_{1} \\
	z_{2} 
	\end{pmatrix}
	= \sqrt{r}	\begin{pmatrix}
		\cos\frac{\theta}{2} e^{\frac{i(\gamma+\phi)}{2}} \\
		\sin\frac{\theta}{2} e^{\frac{i(\gamma-\phi)}{2}}
		\end{pmatrix},
	  \label{t1}
	\end{equation}
which is nothing but  the $SU(2)$ coadjoint orbits in terms of local coordinates and the coordinates $(\sqrt{r},\theta, \phi, \gamma)$  are spherical coordinates on $\mathcal{R}^{4}$ with $r\in [0,\infty)$, $\theta\in[0,\pi]$, $\phi\in[0,2\pi)$ and $\gamma\in[0,2\pi)$ \cite{35}.

Next we introduce the Hilbert space $\mathcal{L}^{2}(\mathcal{C}^{2})$ of square integrable functions on ${\cal C}^2$  with inner product
\begin{equation}
\label{c2inner}
(\phi,\psi)=\int \frac{dz_{\alpha}d\bar{z}_{\alpha}}{\pi^{2}}~ \phi^{*}(z_{\alpha},\bar{z}_{\alpha} ) \psi(z_{\alpha},\bar{z}_{\alpha})
\end{equation}
and the finite norm requirement $(\psi,\psi)<\infty$.    On this space we can now construct a pair of bosonic annihilation and creation operators \cite{37++} as follows

\begin{equation}
    \hat{a}_{\alpha}=z_{\alpha}+\frac{1}{2}\partial_{\bar{z}_{\alpha}};~ ~~\hat{b}_{\alpha}=z_{\alpha}-\frac{1}{2}\partial_{\bar{z}_{\alpha}},
\end{equation}
\begin{equation}
    \hat{a}^\dagger_{\alpha}=\bar{z}_{\alpha}-\frac{1}{2}\partial_{z_{\alpha}};~ ~~\hat{b}^\dagger_{\alpha}=\bar{z}_{\alpha}+\frac{1}{2}\partial_{z_{\alpha}},
\end{equation}
that satisfy the commutation relations (note the unconventional sign)
\begin{equation}
\label{commrel}
    [\hat{a}_{\alpha}, \hat{a}^{\dagger}_{\beta}]=\delta_{\alpha\beta}, ~~ [\hat{b}_{\alpha}, \hat{b}^{\dagger}_{\beta}]=-\delta_{\alpha\beta}~~~\alpha,\beta=1,2
\end{equation}
with all other commutators vanishing.  Note that the creation and annihilation operators are adjoints with respect to the inner product (\ref{c2inner}). 

Now we can readily  define a 4-component bosonic operator-valued quantum spinor \cite{33,34}  (transforming under the $(\frac{1}{2},0)\oplus(0,\frac{1}{2})$ representation of $SU(2)\otimes SU(2))$ acting on $\mathcal{L}^{2}(\mathcal{C}^{2})$ as

\begin{equation}\label{xi0eta0}
\hat{B} = \begin{pmatrix}
\hat{Z}^{+} \\
\hat{Z}^{-}
\end{pmatrix},
\end{equation}
where $ \hat{Z}^{+}=(\hat{a}_{1}~~\hat{a}_{2})^{T}$ , $\hat{Z}^{-}=(\hat{b}_{1}~~\hat{b}_{2})^{T}$ ,$ \hat{Z}^{+\dagger}=(\hat{a}^\dagger_{1}~~\hat{a}^\dagger_{2})^{T}$  and $\hat{Z}^{-\dagger}=(\hat{b}^\dagger_{1}~~\hat{b}^\dagger_{2})^{T}$are two component spinors that satisfy the commutation relations
\begin{equation}
    [\hat{Z}^{\pm}_{\alpha},\hat{Z}^{\pm \dagger}_{\beta}]=\pm\delta_{\alpha \beta} ;~~\alpha,\beta=1,2,
\end{equation}
where $\alpha, \beta$  are the two component spinor indices (note that $\hat{Z}^{+}$ and $\hat{Z}^{-\dagger}$ commute.  The same applies visa versa).

We now construct the following bi-linear forms
\begin{equation}
\label{bil}
    \hat{\Gamma}^A=\hat{\bar{B}} \Gamma^{A}\hat{B}
\end{equation}
with $\bar{\hat{B}}= \hat{B}^{\dagger}\gamma^{0}$ and  $\Gamma^{A}$ a general $4\times4$ matrix. It is simple to check that these have the property
\begin{equation}
    [\hat{\Gamma}^{A},\hat{\Gamma}^{B}]=\hat{\bar{B}} [\Gamma^{A},\Gamma^B]\hat{B}\equiv \hat{\Gamma}^{[A,B]}.
\end{equation}
Note that for this to hold the unconventional sign in (\ref{commrel}) requires $\bar{\hat{B}}$ rather than just ${\hat{B}^\dagger}$ in the definition of the bi-linear forms (\ref{bil}).  This property has the important consequence that the operators $\hat{\Gamma}^A$ satisfy the same algebra as the $4\times 4$ matrices $\Gamma^A$.  Since the Lorentz algebra has a well-known representation in terms of the $4\times 4$ $\gamma$ matrices, we can now readily construct a representation of the Lorentz algebra on $\mathcal{L}^{2}(\mathcal{C}^{2})$.   Similarly, as we know how the $\gamma$ matrices transform under the Lorentz algebra, we can construct tensor operators of the Lorentz algebra on $\mathcal{L}^{2}(\mathcal{C}^{2})$.  Recall that the $\gamma$ matrices are explicitly given by ($\sigma_i$ denotes Pauli spin matrices and $I$ the $2\times 2$ identity matrix)
\begin{equation}\label{gama1}
\gamma^0=\left(
\begin{array}{cc}
I & 0 \\
0 & -I
\end{array}\right), \quad
\gamma^i=\left(
\begin{array}{cc}
0 & \sigma_i \\
-\sigma_i & 0
\end{array}\right),
\quad
\gamma^5=\left(
\begin{array}{cc}
0 & I \\
I & 0
\end{array}\right),
\end{equation}
and that they satisfy the anti-commutation relations $[\gamma^{\mu},\gamma^{\nu}]_{+}=2\eta^{\mu\nu} \mathbb{I};$ ~~$\mu,\nu=0,1,2,3$ with $\eta^{\mu\nu}=(1,-1,-1,-1)$ the Minkowski metric.  From these we can construct a  basis of the Clifford algebra as $\Gamma^{A}=\{ \mathbb{I}, \gamma^{\mu},\sigma^{\mu\nu}, \gamma^{5}, \gamma^{5}\gamma^{\mu}\}$ with $\sigma^{\mu\nu}=\frac{i}{2}[\gamma^{\mu},\gamma^{\nu}]$, which also close on the Lorentz algebra.

We can now construct the following  bi-linear  operators:
\begin{equation}
  \hat{S}=\frac{1}{2}\hat{\bar{B}}\hat{B},~~\hat{D}=-\frac{i}{2}\hat{\bar{B}}\gamma^{5}\hat{B}, ~~ \hat{V}^{\mu}= \frac{1}{2}\hat{\bar{B}}\gamma^{\mu}\hat{B};~~~ \hat{A}^{\mu}= \frac{1}{2} \hat{\bar{B}}\gamma^{5}\gamma^{\mu}\hat{B},
\end{equation}
as well as the two indices anti-symmetric objects
\begin{equation}
    \hat{M}^{\mu\nu}= \frac{1}{2}\hat{\bar{B}}\sigma^{\mu\nu}\hat{B}.
\end{equation}
From the discussion above it is clear that the latter satisfy the usual $so(1,3)$ (Lorentz) algebra:  \begin{equation}
     [\hat{M^{\mu\nu}},\hat{M}^{\rho\sigma}]= -i(\eta^{[\mu[\rho}\hat{M}^{\nu]\sigma]}),
 \end{equation}
with  $\eta_{\mu\nu}$ the inverse Minkowski metric.  From the transformation properties of the $\gamma$ matrices under the Lorentz algebra $\sigma^{\mu\nu}=\frac{i}{2}[\gamma^{\mu},\gamma^{\nu}]$ it is also clear that $\hat{V}^{\mu}$  and $\hat{A}^{\mu}$ transform like 4-vectors:
\begin{equation}
 [\hat{V}^{\mu}, \hat{M}^{\rho\nu}]=i(\eta^{\mu\rho} \hat{V}^{\nu}-\eta^{\mu\nu} \hat{V}^{\rho}),
 \end{equation}\begin{equation}
 [\hat{A}^{\mu}, \hat{M}^{\rho\nu}]=(\eta^{\mu\rho} \hat{A}^{\nu}-\eta^{\rho\nu} \hat{A}^{\rho}),
 \end{equation}
 while $\hat{S}$ and $\hat{D}$ are scalars:
\begin{equation}
     [\hat{S},\hat{M}^{\mu\nu}]=[\hat{D},\hat{M}^{\mu\nu}]=0.
 \end{equation}
Note that all these operators are Hermitian on $\mathcal{L}^{2}(\mathcal{C}^{2})$ so that this is a manifestly unitary representation of the Lorentz algebra.

The 4-vectors also obey the following commutation  relations
\begin{equation}
     [\hat{V}^{\mu}, \hat{V}^{\nu}]=-i\hat{M}^{\mu\nu}, ~ [\hat{A}^{\mu}, \hat{A}^{\nu}]=i\hat{M}^{\mu\nu}~~  [\hat{V}^{\mu}, \hat{A}^{\nu}]=-i\hat{D}\eta^{\mu\nu},~[\hat{V}^{\mu},\hat{D}]= i\hat{A}^{\mu},~~ [\hat{A}^{\mu},\hat{D}]= i\hat{V}^{\mu}
     \label{li}
 \end{equation}
If we exclude $\hat{S}$ and introduce \begin{equation}
    \hat{M}^{5\mu}=\hat{V}^{\mu},~~~~\hat{M}^{4\mu}=\hat{A}^{\mu}~~~\hat{M}^{54}=\hat{D},
\end{equation}
we note that these 15 operators close on the six-dimensional pseudo-orthogonal algebra with the commutation relations:
\begin{equation}
    [\hat{M}^{ab},\hat{M}^{cd}]=-i(\eta^{ac}\hat{M}^{bd}-\eta^{bc} \hat{M}^{ad}+\eta^{bd}\hat{M}^{ac}-\eta^{ad}\hat{M}^{bc})
    \label{eq}
\end{equation}
where $\eta_{ab}=diag(+,-,-,-,-,+)$, for $ a, b = \mu, 4, 5$.\\ 

This is the $so(2,4)$ (anti) de Sitter algebra, which is the isometry group of this six-dimensional Minkowski space-time in that it preserves a hyperboloid. In fact, this group can also be regarded as a group that encompasses conformal transformations of the (3+1)D flat Minkowski space-time \cite{34+,34++}. 

It is useful to have an explicit differential operator representation of these operators on $\mathcal{L}^{2}(\mathcal{C}^{2})$ :
\begin{equation}
\label{cdiff1}
     \hat{S}=\frac{1}{2}(\bar{z}_{\alpha}\partial_{\bar{z}_{\alpha}}-z_{\alpha}\partial_{z_{\alpha}}-2],~~~\hat{D}=\frac{i}{2}(\bar{z}_{\alpha}\partial_{\bar{z}_{\alpha}}+z_{\alpha}\partial_{z_{\alpha}}+2)
 \end{equation}\begin{equation}\label{cdiff2}
    \hat{V}^{0}=[\bar{z}_{\alpha}z_{\alpha}-\frac{1}{4}\partial_{z_\alpha}\partial_{\bar{z}_\alpha}], ~~~~~~~~~\hat{V}^{i}=(\sigma^{i})_{\alpha\beta}[\bar{z}_{\alpha}z_{\beta}+\frac{1}{4}\partial_{z_{\alpha}}\partial_{\bar{z}_{\beta}}];
 \end{equation}\begin{equation}\label{cdiff3}
    \hat{A}^{0}=-(\bar{z}_{\alpha}z_{\alpha}+\frac{1}{4}\partial_{z_{\alpha}}\partial_{\bar{z}_{\alpha}}),~~~~~~\hat{A}^{i}= -(\sigma^{i})_{\alpha\beta}[\bar{z}_{\alpha}z_{\beta}-\frac{1}{4}\partial_{z_{\alpha}\partial_{\bar{z}_{\beta}}}]
\end{equation}
and the Lorentz generators can be written as
\begin{equation}
\label{clordiff1}
\hat{L}_{i}=\frac{1}{2}\epsilon_{ijk}\hat{M}^{jk}= \frac{1}{2}(\sigma_{i})^{\alpha\beta}[\hat{a}^{\dagger}_{\alpha}\hat{a}_{\beta}-\hat{b}^{\dagger}_{\alpha}\hat{b}_{\beta}]=\frac{1}{2}\sigma^{\alpha\beta}_{i} [\bar{z}_{\alpha}\partial_{\bar{z}_{\beta}}-z_{\beta}\partial_{z_{\alpha}}],
\end{equation}
and
\begin{equation}
\label{clordiff2}
\hat{K}_{i}=\hat{M}_{oi}=\frac{i}{2}(\sigma_{i})^{\alpha\beta}[\hat{b}^{\dagger}_{\alpha}\hat{a}_{\beta}-\hat{a}^{\dagger}_{\alpha}\hat{b}_{\beta}]=\frac{i}{2}\sigma_{i}^{\alpha\beta}[z_{\beta}\partial_{z_\alpha}+\bar{z}_{\alpha}\partial_{\bar{z}_{\beta}}],
\end{equation}
where the generators ( $\vec{L}$ and $\vec{K}$) obey the usual $so(1,3)$ algebra:
\begin{equation}
    [\hat{L}_{i},\hat{L}_{j}]=i\epsilon_{ijk} \hat{L}_{k},~ [\hat{K}_i,\hat{K}_{j}]=-i\epsilon_{ijk}\hat{L}_{k}, ~[\hat{L}_{i},\hat{K}_{j}]=i\epsilon_{ijk}\hat{K}_{k}.
\end{equation}
We can introduce a Casimir relation within the framework of $SO(2,4)$ as follows
\begin{equation}
   \hat{C}_2\equiv -\frac{1}{2}\hat{M}_{ab}\hat{M}^{ab}=-\left( \hat{V}_{\mu}\hat{V}^{\mu}-\hat{A}_{\mu}\hat{A}^{\mu}-\hat{D}^{2}+\frac{1}{2}\hat{M}_{\mu\nu}\hat{M}^{\mu\nu}\right),
    \label{ca}
\end{equation}
where we have used the fact that $D^{2}=(\hat{M}^{54})^{2},~\hat{A}_{\mu}=-\hat{M}_{4\mu}$, and ~$\hat{X}_{\mu}=\hat{M}_{5\mu}$.  It is worth noting that the operator $\hat{S}$ plays a significant role. It demonstrates commutativity with all other operators, making it another Casimir operator of $SO(2,4)$ and can be used to label the representations of $SO(2,4)$ carried by $\mathcal{L}^{2}(\mathcal{C}^{2})$.  This will play an important role in what follows.

\subsection{Lorentz covariant deformed Heisenberg-Poincare algebra  }

Similarly to the usual covariant construction in which coordinates and momentum transform as 4-vectors, it is not only natural but essential for covariance that we identify the two 4-vectors constructed above as position and momentum. Of course, these operators do not satisfy the standard Heisenberg algebra, which implies that it is necessary to deform the Heisenberg algebra in order to maintain Lorentz covariance.  Note that our perspective here is that Lorentz covariance is more fundamental than the Heisenberg algebra. We note, though, that there is an ambiguity in this identification, as any linear combination of the two 4-vectors is again a 4-vector and can be identified with position and momentum.  This therefore allows for different deformations of the Heisenberg algebra, depending on the choice of position and momentum as a four-vector. We need to consider two cases, as outlined below:

\subsubsection{Case 1: Commuting coordinates and momenta}

First we introduce
\begin{equation}
    \hat{V}^{\mu}_{\pm}=\frac{1}{2}(\hat{V}^{\mu}\pm\hat{A}^{\mu})
\end{equation}
which satisfy the following commutation relations:
\begin{equation}
    [\hat{V}^{\mu}_{+},\hat{V}^{\nu}_{+}]=0; ~~~ [\hat{V}^{\mu}_{-},\hat{V}^{\nu}_{-}]=0, ~~~~ [\hat{V}^{\mu}_{-},\hat{V}^{\nu}_{+}]=-\frac{i}{2}(\eta^{\mu\nu}\hat{D}+\hat{M}^{\mu\nu}).
\end{equation}

It is instructive to introduce the following dimensionfull operators:
\begin{equation}
\label{comdim}
    \hat{X}^{\mu}=\lambda\hat{V}^{\mu}_{-};~~~~~~~\hat{P}^{\mu}=\frac{2\hbar}{\lambda^{'}}\hat{V}^{\mu}_{+},~~~~\hat{J}^{\mu\nu}=\hbar\hat{M}^{\mu\nu}.
\end{equation}
Here, $\lambda$ represents the length scale linked to the inverse of the local curvature within the curved momentum space, while $\lambda^{'}$ pertains to the length scale associated with the inverse of local curvature in the coordinate space. These then satisfy the commutation relations
\begin{equation}
    [\hat{X}^{\mu},\hat{X}^{\nu}]=0; ~~~ [\hat{P}^{\mu},\hat{P}^{\nu}]=0, ~~~~ [\hat{X}^{\mu},\hat{P}^{\nu}]=-i\hbar\hat{\cal{D}}\eta^{\mu\nu}-i\frac{\lambda}{\lambda^{'}}\hat{J}^{\mu\nu}
    \label{k}.
\end{equation}
where $\hat{\cal{D}}=\frac{\lambda}{\lambda^{'}}\hat{D}$.\\

We note that the coordinates and momenta are commutative as usual, but that the position-momentum commutators are deformed.  Intriguingly, this algebra is not closed; therefore, in order to ensure the closure of the algebra, the standard Heisenberg algebra can be extended to a kinematical algebra with the inclusion of the Lorentz generators ($\hat{J}^{\mu\nu}$) and the central charge ($\hat{\cal{D}}$) for the Lorentz algebra. Thus, the additional commutation relations to eqs. (\ref{k}) are
\begin{equation}
[\hat{X}^{\mu},\hat{\cal{D}}]=-i\frac{\lambda}{\lambda^{'}}\hat{X}^{\mu}~~~[\hat{P}^{\mu},\hat{\cal{D}}]=i\frac{\lambda}{\lambda^{'}}\hat{P}^{\mu},
\end{equation}
and
\begin{equation}
 [\hat{X}^{\mu}, \hat{J}^{\rho\nu}]=i\hbar(\eta^{\mu\rho} \hat{X}^{\nu}-\eta^{\rho\nu} \hat{X}^{\mu}),~~~ [\hat{P}^{\mu}, \hat{J}^{\rho\nu}]=i\hbar(\eta^{\mu\rho} \hat{P}^{\nu}-\eta^{\rho\nu} \hat{P}^{\mu}),\;\; [\hat{\cal{D}},\hat{J}^{\mu\nu}]=0.
 \end{equation}

Finally, let us consider the differential operator representation of these operators.  From (\ref{cdiff3}) these read
\begin{eqnarray}
    \hat{P}^\mu&=&\frac{\hbar}{2\lambda^\prime}\tilde\sigma^\mu_{\alpha\beta}\partial_{z_{\alpha}}\partial_{\bar{z}_{\beta}},\nonumber\\
     \hat{X}^\mu&=&\lambda\sigma^\mu_{\alpha\beta}{\bar z}_{\alpha}z_{\beta}.
\end{eqnarray}
where $\sigma^\mu=\{\mathbb{I},\sigma^i\}$ and $\tilde\sigma^\mu=\{-\mathbb{I},\sigma^i\}$.  It is clear that the momentum operators can be simultaneously diagonalized and that they have continuous spectra.  We also note 
\begin{equation}
    \hat{P}^\mu\hat{P}_\mu=\hat{X}^\mu\hat{X}_\mu=0.
\end{equation}
This situation therefore apparently describes lightcone physics with zero mass.

\subsubsection{Case 2: Non-commuting coordinates and momenta}

In contrast to case 1 above, we can also generate non-commuting coordinates and momenta by identifying them with 4-vectors that are a linear combinations of  the commuting $\hat{X}^\mu$ and $\hat{P}^\mu$ above (the analogue of a Bopp-shift)
         \begin{eqnarray}
         \label{ncomdim}
             \tilde{X}^\mu&=&[c_1\hat{X}^\mu+(\frac{\lambda\lambda^{'}}{2\hbar})d_1\hat{P}^\mu],\\
              \tilde{P}^\mu&=&[c_2\hat{P}^\mu+(\frac{\hbar}{2\lambda\lambda^{'}})d_2\hat{X}^\mu].
         \end{eqnarray}
where $c_1,c_2,d_1,d_2$  are dimensionless real free parameters.   These are non-commuting coordinates and momenta with
\begin{eqnarray}
&&[\tilde{X}^{\mu},\tilde{X}^{\nu}]=-i\frac{\lambda^{2}}{\hbar}\beta_{1}\hat{J}^{\mu\nu}, \ \ \ [\tilde{P}^{\mu},\tilde{P}^{\nu}]=- i\frac{\hbar}{\lambda^{'2}}\beta_{2}\hat{J}^{\mu\nu}, \nonumber \\
&&[\tilde{X}^{\mu},\tilde{P}^{\nu}]=-i\hbar\beta_{0} \hat{\cal{D}}\eta^{\mu\nu}-i\frac{\lambda}{\lambda^{'}} \beta_{3}\hat{J}^{\mu\nu} , \ \ \  [\tilde{X}^{\mu},\hat{\cal{D}}]=i\frac{1}{\beta_{0}}[\frac{\lambda^{2}}{\hbar}\beta_{1}\tilde{P}^{\mu}-\beta_{3}\frac{\lambda}{\lambda^{'}}\tilde{X}^{\mu}], \nonumber \\
&& [\tilde{P}^{\mu},\hat{\cal{D}}]=-i\frac{1}{\beta_{0}}[\frac{\hbar}{\lambda^{'2}}\beta_{2}\tilde{X}^{\mu}-\beta_{3}\frac{\lambda}{\lambda^{'}}\tilde{P}^{\mu}], \nonumber \\
&& [\tilde{X}^{\mu}, \hat{J}^{\rho\nu}]=i\hbar(\eta^{\mu\rho} \tilde{X}^{\nu}-\eta^{\rho\nu} \tilde{X}^{\mu}), \nonumber \\
&&[\tilde{P}^{\mu}, \hat{J}^{\rho\nu}]=i\hbar(\eta^{\mu\rho} \tilde{P}^{\nu}-\eta^{\rho\nu} \tilde{P}^{\mu}), \nonumber \\
&&[\hat{J^{\mu\nu}}\hat{J}^{\rho\sigma}]=-i\hbar(\eta^{\mu\rho}\hat{J}^{\nu\sigma}-\eta^{\nu\rho} \hat{J}^{\mu\sigma}+\eta^{\nu\sigma}\hat{J}^{\mu\rho}-\eta^{\mu\sigma}\hat{J}^{\nu\rho}), \nonumber \\
&&[\hat{\cal{D}},\hat{J}^{\mu\nu}]=0.
\label{algebra}
\end{eqnarray}
where $\beta_{1}=c_{1}d_{1}$, $\beta_{2}=d_{2}c_{2}$, $\beta_{3}=\frac{1}{4}(4c_{1}c_{2}+d_{1}d_{2})$, and $\beta_{0}= \frac{1}{4}(4c_{1}c_{2}-d_{1}d_{2})$.  We also require $\beta_0$ to be non-zero and finite. The latter requirement is needed to avoid a linear dependence between $\tilde{X}^\mu$ and $\tilde{P}^\mu$.  This algebra is nothing but a general deformed Heisenberg Poincare relativistic covariant algebra involving four parameters, $\beta_{1},\beta_{2},\beta_{3}$, and $\beta_{0}$, satisfying $\beta^{2}_{3}-\beta^{2}_{0}=\beta_{1}\beta_{2}$. A four-parameter class of closed algebras is found, but they all exhibit the same stability \cite{36,37,38,39}, as defined by the fact that the isomorphic set of algebras represented by $so(2,4)$ cannot be interpreted as resulting from the Inonu-Wigner contraction \cite{40} of some other group. There are two $so(1, 4)$ sub-algebras for coordinate operators $\hat{\tilde{X}}^{\mu}$ and momentum operators  $\hat{\tilde{P}}^{\mu}$, respectively, with a common $so(1, 3)$ Lorentz algebra. It's evident from (\ref{algebra}) that when we make a specific parameter choice, setting $\beta_{0}=\beta_{1}=-\beta_{2}=1$ and $\beta_{3}=0$, the standard algebra for the AdS version of the Yang model \cite{41,42, 43,44} is effectively reproduced. This standard algebra maintains Lorentz and translational symmetries within fuzzy space-time due to non-commutativity and assumes a background manifold with constant curvature. This results in a duality between the curvature of position and momentum spaces.

These read as differential operators, using (\ref{cdiff3})
\begin{eqnarray}
\label{posmom}
    \tilde{X}^\mu&=&\lambda\left(c_1\sigma^\mu_{\alpha\beta}{\bar z}_{\alpha}z_{\beta}+\frac{d_1}{4}\tilde\sigma^\mu_{\alpha\beta}\partial_{z_{\alpha}}\partial_{\bar{z}_{\beta}}\right),\nonumber\\
     \tilde{P}^\mu&=&\frac{\hbar}{2\lambda^\prime}\left(d_2\sigma^\mu_{\alpha\beta}{\bar z}_{\alpha}z_{\beta}+c_2\tilde\sigma^\mu_{\alpha\beta}\partial_{z_{\alpha}}\partial_{\bar{z}_{\beta}}\right).
\end{eqnarray}
In this case we have
\begin{equation}
       \tilde{P}^\mu\tilde{P}_\mu=\frac{\beta_2\hbar^2}{\beta_1{\lambda^2\lambda^\prime}^2}\tilde{X}^\mu\tilde{X}_\mu=\frac{\beta_2\hbar^2}{{\lambda^\prime}^2}\left(\hat{S}^2+\hat{D}^2+2\hat{S}\right)\equiv \tilde{M}^2v^2
       \label{eq}
\end{equation}
where we have introduced the mass operator $\tilde{M}$ and $v$ is some velocity scale for which the speed of light would be a natural choice.  We do not discuss the spectral properties of these operators further but only note that the spectra of these operators change from discrete to continuous, depending on the signs of $\beta_1$ and $\beta_2$.

\subsection{Cartesian coordinate representation}

Now comes the most important step in this paper, which is the construction of a representation of the (anti) de Sitter and subsequently Lorentz algebra on $\mathcal{L}^{2}(\mathcal{R}^{3})$.   To do this we transform from the complex coordinates in (\ref{t1}) to Cartesian coordinates and the coordinate $\gamma$, i.e.  $(x_1,x_2,x_3,\gamma)$.  Keeping in mind that the Lorentz scalar  $\hat{S}$ commutes with all the generators of the $so(2,4)$ algebra constructed above on $\mathcal{L}^{2}(\mathcal{C}^{2})$, we note that we can use it to construct subspaces of $\mathcal{L}^{2}(\mathcal{C}^{2})$ that are invariant under the action of these operators and hence will also carry a representation of $so(2,4)$.  It is a straightforward but lengthy exercise to compute the Cartesian coordinate representation of these operators from the complex differential operator representation (\ref{cdiff1})-(\ref{clordiff2}) on $\mathcal{L}^{2}(\mathcal{C}^{2})$ using (\ref{t1}).  For $\hat{S}$, we find
\begin{equation}
    \hat{S}=\frac{i}{2}\partial_\gamma-1.
\end{equation}
From this, we note that the subspace of functions of $\mathcal{L}^{2}(\mathcal{C}^{2})$ that depend only on $(x_1,x_2,x_3)$ is the subspace on which $\hat{S}$ takes the value $-1$. We denote these physical states as $\psi_{\rm ph}(z,\bar{z})$. This space of functions is actually not $\mathcal{L}^{2}(\mathcal{R}^{3})$, but rather $\mathcal{L}^{2}(\mathcal{R}^{3},\frac{1}{r})$ as this is the inner product induced on it by the inner product of $\mathcal{L}^{2}(\mathcal{C}^{2})$, as we discuss below. 

One can start by not imposing the constraint $\hat{S}=-1$, but construct the Cartesian coordinate representation of the Lorentz and (anti) de Sitter algebras on the full $\mathcal{L}^{2}(\mathcal{C}^{2})$.   As we do not need the explicit form of these operators, we do not list them here, but for completeness, they are listed in Appendix A.  For our present purposes, it suffices to note that the generic form of these operators is
\begin{equation}
O(a,b,c)=a(x_i,\partial_i)+b(x_i,\partial_i)\partial_\gamma+c(x_i,\partial_i)\partial_\gamma^2
\end{equation}
so that $O(a,b,c)\psi_{\rm ph}(z,\bar{z})=a(x_i,\partial_i)\psi_{\rm ph}(z,\bar{z})$.   We also note that 
\begin{eqnarray}
[O(a,b,c),O(a^\prime,b^\prime,c^\prime)]\psi_{\rm ph}(z,\bar{z})=[a(x_i,\partial_i),a^\prime(x_i,\partial_i)]\psi_{\rm ph}(z,\bar{z})
\end{eqnarray}
so that the operators $a(x_i,\partial_i)$ must satisfy the same algebra on the physical subspace as the operators on the full space.   This implies that the operators projected on the physical subspace must themselves provide a representation of the Lorentz algebra and corresponding 4-vectors.  We can therefore drop all the terms containing $\gamma$ derivatives and focus on the physical subspace only.  For these we find (see Appendix A)
\begin{eqnarray}
\label{3rrepnh}
\hat{L}_k&=&-i\epsilon_{kj\ell}x_j\partial_\ell, \nonumber\\
\hat{K}_k&=& -i r\partial_k,\nonumber\\
   \hat{V}^0_+&=&-\frac{r}{4}\nabla^2, \nonumber\\
  \hat{V}^i_+&=& \frac{1}{4}\left(2\partial_i-x_i\nabla^2+2\left(x\cdot\nabla\right)\partial_i\right),\nonumber\\
     \hat{V}^\mu_-&=&(r,x,y,z),\nonumber\\
    \hat{D}&=&i\left(x\cdot\nabla+1\right).
\end{eqnarray}
It will be noted that these operators are not hermitian with respect to the normal inner product on  $\mathcal{L}^{2}(\mathcal{R}^{3})$, which may look like an inconsistency given that they are hermitian on $\mathcal{L}^{2}(\mathcal{C}^{2})$.  This is, however, clarified when one computes from (\ref{c2inner})  the inner product induced on  $\mathcal{L}^{2}(\mathcal{R}^{3})$:
\begin{eqnarray}
\label{r3inner}
(\phi_{\rm ph},\psi_{\rm ph})&=&\int \frac{dz_{\alpha}d\bar{z}_{\alpha}}{\pi^{2}}~ \phi^{*}_{\rm ph}(z_{\alpha},\bar{z}_{\alpha} ) \psi_{\rm ph}(z_{\alpha},\bar{z}_{\alpha})\nonumber\\
&=&\int \frac{d^3x}{r\pi} \phi^{*}_{\rm ph}(x_i ) \psi_{\rm ph}(x_i)
\end{eqnarray}
where the $\frac{1}{r}$ factor arises from the Jacobian $J=\frac{1}{2r}$.   The operators above are therefore hermitian with respect to this inner product on $\mathcal{L}^{2}(\mathcal{R}^{3})$ and we have actually arrived at an unitary representation of the (anti) de Sitter algebra on $\mathcal{L}^{2}(\mathcal{R}^{3},\frac{1}{r})$.   Finally we can transform to $\mathcal{L}^{2}(\mathcal{R}^{3})$ by performing the similarity transformation $\psi_{\rm ph}(x_i)\rightarrow \tilde{\psi}_{\rm ph}(x_i)=\frac{1}{\sqrt{r}}\psi_{\rm ph}(x_i)$ and $\hat{O}\rightarrow \hat{\tilde{O}}=\frac{1}{\sqrt{r}}\hat{O}\sqrt{r}$ where $\hat{O}$ represents any of the operators above.  Obviously this yields a completely isomorphic algebra, but note that $\tilde{\psi}_{\rm ph}(x_i)$ must now be square integrable.

It is immediately clear that  $\hat{\tilde{S}}=\hat{S }$.  For the other operators we find
\begin{eqnarray}
\label{3rreph}
\hat{\tilde{L}}_k&=&\hat{L}_k=-i\epsilon_{kj\ell}x_j\partial_\ell, \nonumber\\
\hat{\tilde{K}}_k&=& -i \left(r\partial_k+\frac{x_k}{2r}\right),\nonumber\\
   \hat{\tilde{V}}^0_+&=&-\frac{r}{4}\nabla^2-\frac{1}{4r}x\cdot\nabla-\frac{3}{16r},\nonumber \\
    \hat{\tilde{V}}^i_+&=&\frac{1}{4}\left(3\partial_i-x_i\left(\nabla^2+\frac{1}{4r^2}\right)+2\left(x\cdot\nabla\right)\partial_i\right),\nonumber\\
     \hat{V}^\mu_-&=&(r,x,y,z),\nonumber\\
    \hat{\tilde{D}}&=&i\left(x\cdot\nabla+\frac{3}{2}\right).
\end{eqnarray}
One can also verify the following relations:
\begin{eqnarray}
\label{quad}
     &&\hat{\tilde{V}}^\mu_+ \hat{\tilde{V}}_{+\mu}= \hat{\tilde{V}}^\mu_- \hat{\tilde{V}}_{-\mu}=0,\nonumber\\
      &&\hat{\tilde{V}}^\mu_+ \hat{\tilde{V}}_{-\mu}+ \hat{\tilde{V}}^\mu_- \hat{\tilde{V}}_{+\mu}=\hat{\tilde{D}}^2-1.
\end{eqnarray}
Using this in ( \ref{ca}), we can write the second order Casimir as
\begin{equation}
\label{cas}
    \hat{C}_2=\hat{\tilde{K}}^2-\hat{\tilde{D}}^2-\hat{\tilde{L}}^2+2=3.
\end{equation}
This is quite a remarkable result as it implies that we have found a unitary representation of $so(2,4)$ and the Lorentz algebra on $\mathcal{L}^{2}(\mathcal{R}^{3})$.  The fact that the $so(2,4)$ Casimir is a constant also implies that this representation is irreducible.  Note that time is completely absent in this representation.  The question that faces us now is what the implications of the existence of this representation are.  If we take Lorentz covariance as a principle that should guide our construction of physical theories, we must accept that the physical laws should be covariant under these transformations.   In particular this may imply that the Heisenberg has to be deformed to comply with Lorentz covariance.

In this regard, a first check we should carry out is if the inner product (\ref{r3inner})  is Lorentz invariant.  This is clearly the case for rotations.  Under boosts, we note from the discussion above on hermiticity that the boost operator $\hat{B}(a)=e^{ia\cdot \hat{K}}=e^{ra\cdot\nabla}$   is unitary with respect to this inner product so that it is also invariant under boosts.  We can calculate this also more explicitly by noting that due to rotation invariance, we can, without loss of generality, choose the $x_1$-direction as the boost direction, i.e. $\hat{B}(\phi)=e^{\phi r\partial_1}$.  Applying this transformation to the wave-functions we have
\begin{equation}
    \hat{B}(\phi)\psi_{\rm ph}(x_i)=\psi_{\rm ph}(x_i^\prime)
\end{equation}
with 
\begin{equation}
    x_1^\prime=x_1\cosh\phi+r\sinh\phi,\;x^\prime_2=x_2,\;x^\prime_3=x_3.
\end{equation}
Changing variables to $x_i^\prime$ in the integral and computing the Jacobian we find $\frac{d^3x}{r}=\frac{d^3x^\prime}{r^\prime}$ and we conclude
\begin{equation}
(B(\phi)\phi_{\rm ph},B(\phi)\psi_{\rm ph})=(\phi_{\rm ph},\psi_{\rm ph}).
\end{equation}
This result can also be obtained by noting
\begin{equation}
    \hat{B}^\dagger(\phi) \frac{1}{r} B(\phi)=\frac{1}{r}
\end{equation}
where conjugation is taken w.r.t. the normal inner product on $\mathcal{L}^{2}(\mathcal{R}^{3})$.  A final point to note is that after the similarity transformation, the unitary boost operator is given by $\tilde{\hat{B}}(a)=e^{ia\cdot \tilde{\hat{K}}}$ and that, as readily can be checked,  we now have the transformation rule for the wave-functions
\begin{equation}
    \tilde{\hat{B}}(\phi)\tilde{\psi}_{\rm ph}(x_i)=\sqrt{\frac{r^\prime}{r}}\tilde{\psi}_{\rm ph}(x_i^\prime).
\end{equation}
In this case, the $\mathcal{L}^{2}(\mathcal{R}^{3})$ measure is not invariant, but the resulting Jacobian cancels the scaling factors on the wave functions so that the inner product is again Lorentz invariant.

We can again introduce the dimensionful operators of (\ref{comdim}).  Note from (\ref{k}) that position and momentum are commuting operators.  We also note from (\ref{quad}) 
\begin{eqnarray}
\label{dimquad}
     \hat{X}^\mu \hat{X}_{\mu}&=& \hat{P}^\mu \hat{P}_{\mu}=0,\nonumber\\
      \hat{P}^\mu \hat{X}_{\mu}+ \hat{X}^\mu \hat{P}_{\mu}&=&\frac{2\hbar\lambda}{\lambda^\prime}\left(\hat{\tilde{D}}^2-1\right).
\end{eqnarray}
More generally, we can consider the dimensionful non-commuting coordinates and momenta $\tilde{X}^\mu$ and $\tilde{P}^\mu$ of (\ref{ncomdim}). As mentioned before,  they are non-commuting and satisfy the algebra (\ref{algebra}).   We can now compute the Lorentz invariants $\tilde{X}^\mu \tilde{X}_\mu$, $\tilde{P}^\mu \tilde{P}_\mu$ and $\tilde{P}^\mu \tilde{X}_\mu+\tilde{X}^\mu \tilde{P}_\mu$ using (\ref{quad}) to find 
\begin{eqnarray}
\label{relations}
           \tilde{X}^\mu \tilde{X}_\mu&=&\beta_1\lambda^2\left(\hat{\tilde{D}}^2-1\right),\\
           \tilde{P}^\mu \tilde{P}_\mu&=&\beta_2\left(\frac{\hbar}{\lambda^\prime}\right)^2\left(\hat{\tilde{D}}^2-1\right)\equiv \tilde{M}^2v^2\nonumber\\
           \tilde{P}^\mu \tilde{X}_{\mu}+ \tilde{X}^\mu \tilde{P}_{\mu}&=&\frac{2\beta_3\hbar\lambda}{\lambda^\prime}\left(\hat{\tilde{D}}^2-1\right).
\end{eqnarray}

If we make the usual identification of $\tilde{E}=\tilde{P}^0$ as the energy (Hamiltonian), we obtain the usual relativistic dispersion relation
\begin{equation}
    \tilde{E}=\pm\sqrt{\hat{M}^2v^2+{\vec{\tilde{P}}}\cdot\vec{\tilde{P}}}
\end{equation}

\subsection{The Heisenberg algebra as an In\"onu–Wigner contraction of the deformed Heisenberg Poincare algebra}

We now demonstrate that, on a certain sector of the Hilbert space, the deformed Heisenberg-Poincaré algebra introduced above can be contracted to the conventional Heisenberg-Poincaré algebra associated with Minkowski space-time.  To do this, we consider a fixed mass sector, i.e., $\tilde{M}^2|m\rangle=m^2|m\rangle$  in the limit of large $\lambda^\prime$, i.e., small curvature in coordinate space, and for $\beta_2>0$.  From (\ref{relations}) we then have 
\begin{equation}
\hat{\tilde{D}}|m\rangle=\sqrt{1+\frac{{\lambda^\prime}^2m^2v^2}{\hbar^2\beta_2}}|m\rangle\approx \frac{{\lambda^\prime}mv}{\hbar\sqrt{\beta_2}}|m\rangle.
\end{equation}
From (\ref{algebra}), we note that on this sector the position-momentum commutator assumes the form
\begin{equation}
    [\tilde{X}^{\mu},\tilde{P}^{\nu}]=-i\frac{\lambda mv\beta_{0}}{\sqrt{\beta_2}} \hat{\mathbb{I}}_m\eta^{\mu\nu}-i\frac{\lambda}{\lambda^{'}} \beta_{3}\hat{J}^{\mu\nu} 
\end{equation}
with $\mathbb{I}_m$ the identity (projector) on this sector.  This will reduce to the standard position-momentum commutator on this sector if we set $\frac{\lambda mv\beta_{0}}{\sqrt{\beta_2}}=\hbar$.   We then have for small $\lambda$, i.e. small curvature in momentum space,
\begin{equation}
    \beta_3=\sqrt{\beta_2}\sqrt{\beta_1+\frac{\hbar^2}{\lambda^2m^2v^2}}\approx\frac{\hbar\sqrt{\beta_2}}{\lambda mv}.
\end{equation}
This turns (\ref{algebra}) into
\begin{eqnarray}
&&[\tilde{X}^{\mu},\tilde{X}^{\nu}]=-i\frac{\lambda^{2}}{\hbar}\beta_{1}\hat{J}^{\mu\nu}, \ \ \ [\tilde{P}^{\mu},\tilde{P}^{\nu}]=- i\frac{\hbar}{\lambda^{'2}}\beta_{2}\hat{J}^{\mu\nu}, \nonumber \\
&&[\tilde{X}^{\mu},\tilde{P}^{\nu}]=-i \hat{\mathbb{I}}_m\eta^{\mu\nu}-i\frac{\hbar\sqrt{\beta_2}}{\lambda^\prime mv}\hat{J}^{\mu\nu}  , \ \ \  [\tilde{X}^{\mu},\hat{\cal{D}}]=i\frac{\lambda mv}{\hbar\sqrt{\beta_2}}[\frac{\lambda^{2}}{\hbar}\beta_{1}\tilde{P}^{\mu}-\frac{\hbar\sqrt{\beta_2}}{ mv\lambda^{'}}\tilde{X}^{\mu}], \nonumber \\
&& [\tilde{P}^{\mu},\hat{\cal{D}}]=-i\frac{\lambda mv}{\hbar\sqrt{\beta_2}}[\frac{\hbar}{\lambda^{'2}}\beta_{2}\tilde{X}^{\mu}-\frac{\hbar\sqrt{\beta_2}}{ mv\lambda^{'}}\tilde{P}^{\mu}], \nonumber \\
&& [\tilde{X}^{\mu}, \hat{J}^{\rho\nu}]=i\hbar(\eta^{\mu\rho} \tilde{X}^{\nu}-\eta^{\rho\nu} \tilde{X}^{\mu}), \nonumber \\
&&[\tilde{P}^{\mu}, \hat{J}^{\rho\nu}]=i\hbar(\eta^{\mu\rho} \tilde{P}^{\nu}-\eta^{\rho\nu} \tilde{P}^{\mu}), \nonumber \\
&&[\hat{J^{\mu\nu}}\hat{J}^{\rho\sigma}]= -i\hbar(\eta^{\mu\rho}\hat{J}^{\nu\sigma}-\eta^{\nu\rho} \hat{J}^{\mu\sigma}+\eta^{\nu\sigma}\hat{J}^{\mu\rho}-\eta^{\mu\sigma}\hat{J}^{\nu\rho}), \nonumber \\
&&[\hat{\cal{D}},\hat{J}^{\mu\nu}]=0.
\label{algebracont}
\end{eqnarray}
Finally, taking the limits $\lambda\rightarrow 0$ and $\lambda^\prime\rightarrow\infty$, we obtain the standard Poincare algebra on this sector
\begin{eqnarray}
&&[\tilde{X}^{\mu},\tilde{X}^{\nu}]=0, \ \ \ [\tilde{P}^{\mu},\tilde{P}^{\nu}]=0, \nonumber \\
&&[\tilde{X}^{\mu},\tilde{P}^{\nu}]=-i \hat{\mathbb{I}}_m\eta^{\mu\nu}  , \ \ \  \nonumber \\
&& [\tilde{X}^{\mu}, \hat{J}^{\rho\nu}]=i\hbar(\eta^{\mu\rho} \tilde{X}^{\nu}-\eta^{\rho\nu} \tilde{X}^{\mu}), \nonumber \\
&&[\tilde{P}^{\mu}, \hat{J}^{\rho\nu}]=i\hbar(\eta^{\mu\rho} \tilde{P}^{\nu}-\eta^{\rho\nu} \tilde{P}^{\mu}), \nonumber \\
&&[\hat{J^{\mu\nu}}\hat{J}^{\rho\sigma}]=- i\hbar(\eta^{\mu\rho}\hat{J}^{\nu\sigma}-\eta^{\nu\rho} \hat{J}^{\mu\sigma}+\eta^{\nu\sigma}\hat{J}^{\mu\rho}-\eta^{\mu\sigma}\hat{J}^{\nu\rho}). \nonumber \\
\label{poincare}
\end{eqnarray}
Note that the operator $\hat{\cal{D}}$  contracts to the null operator in this limit, and we therefore do not list its commutators here as they all vanish.  Essentially, this result implies that in flat space and momentum space, the coordinates and momenta are close to the standard Heisenberg algebra, and mass becomes a good quantum number.  Obviously, this means that in the presence of weak curvature, the Heisenberg algebra is a good approximation to the deformed Heisenberg-Poincare algebra.

\section{Covariant quantum dynamics}
\label{dynamics}

So far, our focus has centered on foundational aspects at the kinematical level. In this section, we elevate our exploration to introduce Lorentz covariant dynamics. Consider the representation of the Lorentz algebra on $\mathcal{L}^{2}(\mathcal{R}^{3})$ with generators $\hat{\tilde{L}}_{i}(t=0)$ and $\hat{\tilde{K}}_{i}(t=0)$ which leave the physical subspace
\begin{equation}
\frac{\partial}{\partial \gamma}\psi_{ph}(z,\bar{z})=0
\end{equation}
invariant.  We now define time evolution in the physical subspace as a one-parameter family of states:
\begin{equation}
\label{time}
\psi_{ph}(x_{1},x_{2},x_{3},t)=e^{-it\hat{H}}\psi_{ph}(x_{1},x_{2},x_{3},t=0)\equiv \hat{U}(t)\psi_{ph}(x_{1},x_{2},x_{3},t=0).
\end{equation}
Here, $\hat{H}$ is for the moment a time-independent Hermitian operator that commutes with the constraint. A natural choice is the energy $\hat{H}=\hat{\tilde{P}}^0$.  This ensures that

\begin{equation}
\frac{\partial}{\partial \gamma}\psi_{ph}(x_{1},x_{2},x_{3},t)=0, \quad \forall t.
\end{equation}
Clearly,

\begin{equation}
[i\partial_{t}-\hat{H}]\psi_{ph}(x_{1},x_{2},x_{3},t)=0.
\label{LO}
\end{equation}

When defining these time-dependent states through the unitary transformation $\hat{U}(t)$, we also need to transform the differential operator representation of the Lorentz generators:

\begin{equation}
\hat{\tilde{L}}_{i}(t)=\hat{U}(t)\hat{\tilde{L}}_{i}\hat{U}^{\dagger}(t), \quad \hat{\tilde{K}}_{i}(t)=\hat{U}(t)\hat{\tilde{K}}_{i}\hat{U}^{\dagger}(t).
\end{equation}
It may be noted that

\begin{equation}
i\hat{U}(t)\partial_{t} \hat{U}^{\dagger}(t)=i\partial_{t}-\hat{H}.
\end{equation}
Clearly,
\begin{equation}
[\hat{L}_{i}(t=0),\frac{\partial}{\partial t}]=0, \quad [\hat{K}_{i}(t=0),\frac{\partial}{\partial t}]=0.
\end{equation}
On applying the unitary transformation, we have

\begin{equation}
[\hat{L}_{i}(t),\frac{\partial}{\partial t}-\hat{H}]=0, \quad [\hat{K}_{i}(t),\frac{\partial}{\partial t}-\hat{H}]=0.
\end{equation}
Thus, $i\partial_{t}-\hat{H}$ is a Lorentz scalar, and consequently, (\ref{LO}) must represent a fully covariant equation of motion.  Note that the Lorentz generators now have, as usual, an explicit time dependence.

The argument above can easily be generalised to time-dependent Hamiltonian by simply replacing the exponential in (\ref{time}) by a path-ordered exponential.  The rest of the argument remains the same.\\

\section{The Fuzzy $\mathcal{R}_{\lambda}^{3}$ representation of the Lorentz and (anti) de Sitter algebras. }
\label{Fuzzy}

In this section we construct an alternative representation of the Lorentz and (anti) de Sitter algebras on fuzzy three dimensional space and explore its connection with the ${\cal L}^2({\cal R}^3)$  construction in the previous section.

\subsection{Quantum mechanics on fuzzy ${\cal R}^3_{\lambda}$}

Three dimensional fuzzy space assumes the fuzzy sphere $su(2)$ commutation relations for coordinates
\begin{equation}
[\hat{x}_i,\hat{x}_j]=2i\lambda\varepsilon_{ijk}\hat{x}_k.
\end{equation}
Here $\lambda$ has units of length and $\varepsilon_{ijk}$ is the standard completely anti-symmetric tensor.  The first step in constructing a quantum theory on this space is to construct a concrete realisation of the fuzzy sphere non-commutative coordinates on some Hilbert space that we'll denote by ${\cal H}_c$ .  The representation we choose for this purpose is the standard Schwinger realisation of $su(2)$.  Thus, ${\cal H}_c$ is a two-mode Fock space on which the coordinates are realised as
\begin{equation}
\hat{x}_i=\lambda a^\dagger_\alpha \sigma^{(i)}_{\alpha\beta} a_\beta.
\end{equation}
Here, a summation over repeated indices is implied. $\alpha,\beta=1,2$, $\sigma^{(i)}_{\alpha\beta}$, $i=1,2,3$ are the Pauli spin matrices, and $a_\alpha^\dagger$, $a_\alpha$ are standard boson creation and annihilation operators.  The radius operator is 
\begin{equation}
\hat{r}^2=\hat{x}_i\hat{x}_i =\lambda^2 \hat{n}(\hat{n}+2),
\end{equation}
with $\hat{n}=a_\alpha^\dagger a_\alpha$ the boson number operator.  Note that the radius operator is also the Casimir of $su(2)$ and commutes with the coordinates.    As a measure of the radius we use 
\begin{equation}
\label{radiusc}
\hat{r}=\lambda (\hat{n}+1),
\end{equation}
which is to leading order in $\lambda$ the square root of $\hat{r}^2$.  Note that this representation contains each $su(2)$ representation, and thus each quantised radius, exactly once and therefore corresponds to a complete single covering of ${\cal R}^3$, commonly referred to as ${\cal R}^3_{\lambda}$ fuzzy space.

The next step is to construct the quantum Hilbert space ${\cal H}_q$ in which the states of the system are to be represented.  In analogy with ${\cal L}^2({\cal R}^3)$ this is now defined as the algebra of operators generated by the coordinates, i.e. the operators acting on ${\cal H}_c$ that commute with $\hat{r}^2$ and have a finite norm with respect to a weighted Hilbert-Schmidt inner product \cite{23}
\begin{eqnarray}
\label{qhilbert}
{\cal H}_q=\left\{\psi: [\psi,\hat{r}^2]=0\ \ {\rm and}\ \ {\rm tr_c}(\psi^\dagger\hat{r}\psi)<\infty\right\}.
\end{eqnarray}
The inner product on ${\cal H}_q$ is
\begin{equation}
\label{innp}
	(\psi|\phi)_q=4\pi\lambda^2{\rm tr}_c(\psi^\dagger\hat{r}\phi)=4\pi\lambda^3{\rm tr}_c(\psi^\dagger\left(\hat{n}+1\right)\phi)
\end{equation}
with the trace taken over ${\cal H}_c$.  We use the standard $|\cdot\rangle$ notation for elements of ${\cal H}_c$ and $|\cdot)$ for elements of ${\cal H}_q$. This choice of the inner product is motivated by the observation that the norm of the operator that projects on the subspace of spheres with radius $r\leq \lambda(N+1)$, with $N$ large, corresponds to the volume of a sphere in three dimensional Euclidean space.

 Quantum observables are identified with self-adjoint operators acting on ${\cal H}_q$. We use capitals to distinguish them from operators acting on ${\cal H}_c$.  These include the coordinates which act through left multiplication as
\begin{equation}
	\hat{X}_i|\psi)=|\hat{x}_i\psi)
\end{equation}
and the angular momentum operators which act adjointly according to
\begin{equation}
	\hat{L}_i|\psi) = |\frac{\hbar}{2\lambda}[\hat{x}_i, \psi])\quad{\rm with}\quad[\hat{L}_i,\hat{L}_j] = i\hbar\varepsilon_{ijk}\hat{L}_k.
\label{eq:nc-angular-ops}
\end{equation}
There is a further important conserved quantity, namely, the operator $\hat\Gamma$, which acts as follows
\begin{equation}
\hat\Gamma|\psi)=|[a^\dagger_\alpha a_\alpha,\psi]).
\end{equation}
From the defining relation (\ref{qhilbert}) it is clear that this operator tests whether a state is actually in ${\cal H}_q$ and we refer to it as the physicality condition.  Physical observables are required to commute with $\hat\Gamma$ in order to leave ${\cal H}_q$ invariant.

\subsection{The representation of the Lorentz and (anti) de Sitter algebras on fuzzy ${\cal R}_{\lambda}^3$}

Let us enlarge ${\cal H}_q$ to the Hilbert space ${\cal H}$ defined as follows:
\begin{eqnarray}
\label{hilbert}
{\cal H}=\left\{\psi: {\rm tr_c}(\psi^\dagger\psi)<\infty\right\},
\end{eqnarray}
i.e. we dropped the physicality condition and adopted the inner product 
\begin{equation}
\label{innp1}
	(\psi|\phi)=4\pi\lambda^2{\rm tr}_c(\psi^\dagger\phi).
\end{equation}
We note that 
\begin{equation}
	(\psi|\phi)=(\psi|\frac{1}{\hat{r}}|\phi)_q,
\end{equation}
reminiscent of (\ref{r3inner}).  

On this space we define the following bosonic operators
\begin{eqnarray}
&&A_{\alpha L}^\dagger|\psi)=|a_\alpha^\dagger\psi),\quad A_{\alpha L}|\psi)=|a_\alpha\psi),\nonumber\\
&&A_{\alpha R}^\dagger|\psi)=|\psi a_\alpha^\dagger),\quad A_{\alpha R}|\psi)=|\psi a_\alpha).
\end{eqnarray}
It is easily checked that these operators are hermitian conjugates with respect to the inner product (\ref{innp1}).  Also note that 
\begin{equation}
[A_{\alpha L},A_{\beta L}^\dagger]=\delta_{\alpha\beta},\quad [A_{\alpha R},A_{\beta R}^\dagger]=-\delta_{\alpha\beta}
\end{equation}
and all other commutators vanish. We now note a complete isomorphism with the construction in Section 2,  under the map
\begin{equation}
     A_{\alpha L}\rightarrow \hat{a}_\alpha,\; A^\dagger_{\alpha L}\rightarrow \hat{a}^\dagger_\alpha,\;  A_{\alpha R}\rightarrow \hat{b}_\alpha,\; A^\dagger_{\alpha R}\rightarrow \hat{b}^\dagger_\alpha.
\end{equation}
We can therefore proceed in exactly the same way as there to construct a representation of the Lorentz and (anti) de Sitter algebras as bilinears of these creation and annihilation operators acting on $\cal{H}$.  It is then easy to establish that $\hat{S}$ and $\hat{\Gamma}$ play the same role, i.e., the restriction to the physical subspace ${\cal H}_q$ is equivalent to the restriction to ${\cal L}^2({\cal R}^3)$.  The rest of the analysis is the same as in Section 2.

\subsection{The link between the ${\cal L}^2({\cal R}^3)$ and fuzzy ${\cal R}^3_{\lambda}$ representations of the Lorentz and (anti) de Sitter algebras}

From the above, it should be clear that we now have two representations of the Lorentz and (anti) de Sitter algebras, one on ${\cal L}^2({\cal R}^3)$ and one on fuzzy ${\cal R}^3_{\lambda}$ that are, by construction, algebraically equivalent.  This equivalence can be made even more explicit by noting the following:  For the noncommutative spatial and momentum sectors of the deformed Heisenberg algebra, given by:

\begin{equation}
  [\tilde{X}_{i},\tilde{X}_{j}]=-i\frac{\lambda^{2}}{\hbar}\beta_{1}\hat{J}_{ij},~~  [\tilde{P}_{i},\tilde{P}_{j}]=-i\frac{\hbar}{\lambda^{'2}}\beta_{2}\hat{J}_{ij},
\end{equation}

where $\hat{J}_{ij}=\epsilon_{ijk}\hat{L}^{k}$ and  $[\hat{L}_{i},\hat{L}_{j}]=i\hbar\epsilon_{ijk}\hat{L}_{k}$ we can define a new basis for the spatial sector of the $so(2,4)$ algebra as:

\begin{equation}
    \hat{x}_{i}=\tilde{X}_{i}+\frac{\lambda\sqrt{\beta_{1}}}{\hbar}\hat{L}_{i},~~~\hat{p}_{i}=\tilde{P}_{i}+\frac{\sqrt{\beta_{2}}}{\lambda^{'}}\hat{L}_{i}.
\end{equation}
Then, the spatial sector of the (anti) de Sitter algebra (\ref{algebra}) has the following structure:
\begin{eqnarray}
[\hat{x}_i, \hat{x}_j] &=&2i\lambda\sqrt{\beta_{1}} \epsilon_{ijk}\hat{x}_{k},  \,\,\,\,\,\,  [\hat{L}_i, \hat{x}_j]= i\hbar\epsilon_{ijk}\hat{x}_{k}, \cr
[\hat{p}_i, \hat{p}_j] &=&  \frac{2\hbar\sqrt{\beta_{2}}}{\lambda^{'}}\epsilon_{ijk} \hat{p}_k,\,\,\,[\hat{x}_{i}, \hat{p}_j]= i\hbar\beta_{0}\hat{D}+i\frac{\lambda}{\lambda^{'}}(\beta_{3}+\sqrt{\beta_{1}\beta_{2}})\epsilon_{ijk}\hat{L}_{k}-i\lambda\sqrt{\beta_{1}}\epsilon_{ijk}\hat{p}_{k}+i\frac{\hbar\sqrt{\beta_{2}}}{\lambda^{'}}\epsilon_{ijk}\hat{x}_{k},\cr
[\hat{L}_{i}, \hat{p}_j]&=&  i\hbar\epsilon_{ijk}\hat{p}_{k},\,\,\,\,\,\,\,\,\,\,\, [\hat{x}_{i},\hat{\mathcal{D}}]=i\frac{1}{\beta_{0}}[\frac{\lambda^{2}}{\hbar}\beta_{1}\hat{p}_{i}-\beta_{3}\frac{\lambda}{\lambda^{'}}\hat{x}_{i}+\frac{\lambda^{2}}{\lambda^{'}\hbar}(\beta_{3}\sqrt{\beta_{1}}-\beta_{1}\sqrt{\beta_{2}})\hat{L}_{i}],\cr
[\hat{p}_{i},\hat{\mathcal{D}}]&=& - i\frac{1}{\beta_{0}}[\frac{\hbar\beta_{2}}{\lambda^{'2}}\hat{x}_{i}-\beta_{3}\frac{\lambda}{\lambda^{'}}\hat{p}_{i}+\frac{\lambda}{\lambda^{'2}}(\beta_{3}\sqrt{\beta_{2}}-\beta_{2}\sqrt{\beta_{1}})\hat{L}_{i}].
\label{finaldesitteralgebra}
\end{eqnarray}
where the form of $\hat{\mathcal{D}}$ is already defined through the Casimir relationship given in equation (\ref{ca}). It is noteworthy that the above structure bears resemblance to a nonstandard closure of the spatial components of the deformed Heisenberg algebra. Indeed, it is crucial to note that the commutation between spatial coordinates, as expressed in equation (\ref{finaldesitteralgebra})
\begin{equation}\label{alg}
[\hat{x}_i,\hat{x}_j] = 2i\lambda\epsilon_{ijk}\hat{x}_k
\end{equation}
illuminates a captivating geometric insight \cite{45}. This commutation relation unveils a three-dimensional space characterized by a set of concentric fuzzy spheres of intersecting radii. This intricate spatial structure emerges prominently when considering specific parameter values, specifically $\beta_{1}=-\beta_{2}=\beta_{0}=1$ and $\beta_{3}=1$, chosen for brevity and clarity. This structure has surfaced as a solution to modified Einstein equations, resembling a black hole solution \cite{46}, UV/IR mixing freedom \cite{48}, and quantum gauge theory \cite{49}. Additionally, it has been independently scrutinized across different quantum mechanical contexts, as referenced, for instance, in \cite{47}.\\

Although the two representations above are algebraically equivalent, they may be non-equivalent as representations.  For this, one has to establish an isometry between ${\cal H}_q$ and ${\cal L}^2({\cal R}^3)$.  This falls outside the scope of the present paper but will be discussed in a follow-up.

\section{ Conclusion and Future Outlook}
\label{Conclusions}

Let us summarise, point by point, the new significant findings of the paper, comparing with existing results found in
the literature.

\begin{itemize}
\item 

In special relativity, space and time are treated as coordinates within a smooth, commutative, flat manifold. This framework provides a unitary space-time representation of the Lorentz algebra, $so(1,3)$, acting on square-integrable functions over the space-time manifold. However, when transitioning to quantum mechanics, there is a tendency to assume the existence of absolute time. This assumption serves as a reference for describing time as an evolutionary parameter at the quantum level.

In the Schrodinger picture, the state vectors initially exhibit no time dependency, and time dependence is introduced to these states through their evolution. Hence, the time ``coordinate" $t$ is simply a parameter susceptible to arbitrary selection and modification by the observer's perspective \cite{47+++}. This prompts the intriguing question:  can we introduce the representation of 
$so(1,3)$ and the concept of Lorentz four vectors in a ``timeless" manner at the quantum level? The answer is affirmative.

In this paper, a unitary representation of the Lorentz and (anti) de Sitter algebras has been constructed on ${\cal L}^2({\cal R}^3)$. Notably, time remains entirely absent within this representation. However, it can be introduced in a fully covariant manner through a parameter associated with a time translation operator. This operator can naturally be selected as the zeroth component of the non-commutative momentum four vector.

\item 
To construct this fully covariant formulation of quantum mechanics, we required a deformation of the Heisenberg phase space algebra together with Poincare algebra; however, it appears naturally, which may affect the entire quantum mechanical framework of
fields and particles. Here, we also observe that the zeroth component of the position four vector does not represent the time coordinate.
In our formulation, Lorentz symmetry holds greater significance than the Heisenberg phase space algebra at the quantum level. Essentially, the Heisenberg algebra emerges as a consequence of the undeformed Lorentz algebra realized on ${\cal L}^2({\cal R}^3)$. This deformed Heisenberg algebra can be categorized into two equivalent forms, where one entails commutative coordinates and momentum. Conversely, the other involves a non-commutative phase space Lie algebra, akin to the one deduced by Mandes \cite{37} and Yang \cite{41}. Notably, our non-commutative deformed Heisenberg-Poincare algebra also shares correspondence with a linear (Lie algebraic) rendition of the non-linear triple special relativity (TSR) phase space algebra proposed by Gilkman and Smolin \cite{48}, following the correct identification of its generators. 

\item 
Moreover, it's crucial to note that the modification in our deformed phase space algebra explicitly signifies an inherent curvature in both coordinate and momentum spaces. This curvature is revealed through the non-commutative structure of phase space variables \cite{49}, wherein the curvature scales in position and momentum spaces are contingent upon the fundamental length scales 
$\lambda$ and $\lambda^{'}$, respectively. Remarkably, this curvature persists even in the absence of conventional sources.

The mass of an individual particle is influenced by the momentum space's curvature scale 
$\lambda^{'}$—an element akin to the cosmological constant in General Relativity (GR)—and the Planck constant ($\hbar$), which characterizes the quantum world, encapsulating the wave nature of particles. We have also demonstrated that our representation of the deformed Heisenberg-Poincaré algebra on 
${\cal L}^2({\cal R}^3)$ can be contracted to the conventional Heisenberg and Poincaré algebra by taking appropriate flat limits within fixed mass sectors of the physical Hilbert space.
Also note that this implies the necessity of curvature scales to achieve this unique representation of the Lorentz algebra in the quantum domain.

In conclusion, unlocking Lorentz covariance at the level of the Schr\"odinger evolution in quantum states leads to the recovery of a deformed Heisenberg-Poincaré algebra. This strongly implies that our deformed Heisenberg-Poincaré algebra is instrumental in delineating physical phenomena at the interface of gravitational and quantum realms (IGQR).\\\\

\item 

We also demonstrated that the same algebraic construction can be made on fuzzy ${\cal R}^3_{\lambda}$, but at this stage, it is not clear whether these constructions are truly equivalent as representations.  This will be explored elsewhere.\\

\end{itemize}

We close this section with some remarks on future developments.  The most pressing issue is how standard, classical (special) relativity emerges from this picture in light of the different role time plays here.  Relativistic effects are well-tested phenomena, and if they cannot be reproduced, at least in some approximation, this may invalidate the construction. Another obvious outstanding issue is the physical content of the current construction.  For this, one would like to explore the spectral properties of the observables in the two constructions, as these have specific physical consequences analysed in \cite{26,27,30,31,32} for fuzzy ${\cal R}^3_{\lambda}$. The applicability of these results in the case of the ${\cal L}^2({\cal R}^3)$ representation hinges critically on the equivalence of these representations and subsequently the spectral properties of the observables.  An important issue that needs to be explored further is the possible impact of this construction on cosmic, galactic, and solar length scales.  In particular, one would like to explore whether it may lead to Modified Newton Dynamics (MOND).  One strongly suspects that this may be the case, as this was already confirmed in \cite {32} for fuzzy ${\cal R}^3_{\lambda}$.  This may also have a bearing on the Hubble tension problem \cite {51,52}.

There are also several other possible experimental routes to explore our proposed top-down approach to Lorentz-covariant quantum mechanics. For example, recently, there have been many phenomenological studies in atomic physics, as some experiments in atomic physics are now sensitive to
small frequency shifts below 1 mHz \cite{53}. With such sensitivity, IGQR effects might be detected at
low energies, especially if small energy shifts have a qualitative impact. Here, such a possibility is analyzed by looking at the effect of the non-commutative algebra on the
spectrum of the quantum Coulomb problem, even if the corrections turn out to be too small to be measurable in the
near future.

Furthermore, within our formulation, non-commutative momentum signals the modification of zero-point energy eigenvalues (see \cite{50} for detail), a factor measurable through precise laboratory experiments involving Planck mass oscillators and superconducting quantum interference devices (SQUIDs). This is evident as SQUIDs are known to conduct supercurrents with a tunable superconducting mass dependent on temperature, potentially providing a laboratory signature \cite{54} of the IGQR effect. 

\section{Acknowledgements}
One of the authors (PN) acknowledges the support from a 
postdoctoral fellowship grants from Stellenbosch University, South Africa.

\section*{Appendix A}

Transforming from the complex variables to the Cartesian coordinates $(x_1,x_2,x_3)$ and $\gamma$ we find for the Lorentz generators
\begin{eqnarray}
\hat{L}_k&=&\frac{ir\alpha_k}{2 \left(x_1^2+x_2^2\right)}\partial_{\gamma }-i\epsilon_{kj\ell}x_j\partial _\ell \\
\hat{K}_k&=& \frac{i\beta_k}{2 \left(x_1^2+x_2^2\right)}\partial_{\gamma }-i r\partial_k
\end{eqnarray}
where $\alpha_k=\left(x_1,-x_2,0\right)$, $\beta_k=\left(x_2x_3,-x_1x_3,0\right)$ and  $r=\sqrt{x_1^2+x_2^2+x_3^2}$. 

Next we consider the Lorentz scalar  $\hat{D}$.  Since this is a Lorentz scalar, it commutes with the Lorentz generators and can be used to label the SO(1,3) representations carried by $\mathcal{L}^{2}(\mathcal{R}^{3})$ .  Note though that this operator does not commute  with all the bilinears constructed above.  We find its Cartesian representation to be
\begin{equation}
    \hat{D}=i\left(x\cdot\nabla+1\right),
\end{equation}
which is just the dilation operator.

For the commutative 4-vector  $\hat{V}^\mu_-$ we find
\begin{equation}
    \hat{V}^\mu_-=(r,x,y,z),
\end{equation}
which justifies the interpretation of these operators as commuting coordinates, and  for the commutative 4-vector $\hat{V}^\mu_+$ 
\begin{eqnarray}
    \hat{V}^0_+&=&-\frac{1}{16 \left(x_1^2+x_2^2\right)}\left(r \partial^{}_{\gamma }-4 x_2 x_3  \partial_1+4 x_1 x_3  \partial_2\right)\partial_{\gamma }-\frac{r}{4}\nabla^2 \nonumber
    \\
    \hat{V}^1_+&=&\frac{1}{16 \left(x_1^2+x_2^2\right)}\left(x_1 \partial_{\gamma }-4 r x_2  \partial_3\right)\partial_{\gamma }+\frac{1}{4}\left(2\partial_1-x_1\nabla^2+2\left(x\cdot\nabla\right)\partial_1\right)\nonumber\\
    \hat{V}^2_+&=&\frac{1}{16 \left(x_1^2+x_2^2\right)}\left(x_2 \partial_{\gamma }+4 r x_1  \partial_3\right)\partial_{\gamma }+\frac{1}{4}\left(2\partial_2-x_2\nabla^2+2\left(x\cdot\nabla\right)\partial_2\right)\nonumber\\
    \hat{V}^3_+&=&\frac{1}{16 \left(x_1^2+x_2^2\right)}\left(-x_3 \partial_{\gamma }+4 r \left(x_2\partial_1-x_1\partial_2\right)\right)\partial_{\gamma }+\frac{1}{4}\left(2\partial_3-x_3\nabla^2+2\left(x\cdot\nabla\right)\partial_3\right)\nonumber\\
\end{eqnarray}

\end{document}